# The Final Chapter In The Saga Of YIG

A. J. Princep[1]*, R. A. Ewings[2], S. Ward[3], S. Tóth[3], C. Dubs[4], D. Prabhakaran[1], A. T. Boothroyd[1]

**The magnetic insulator Yttrium Iron Garnet can be grown with exceptional quality, has a ferrimagnetic transition temperature of nearly 600 K, and is used in microwave and spintronic devices that can operate at room temperature[1]. The most accurate prior measurements of the magnon spectrum date back nearly 40 years, but cover only 3 of the lowest energy modes out of 20 distinct magnon branches[2]. Here we have used time-of-flight inelastic neutron scattering to measure the full magnon spectrum throughout the Brillouin zone. We find that the existing model of the excitation spectrum, well known from an earlier work titled "The Saga of YIG" [3], fails to describe the optical magnon modes. Using a very general spin Hamiltonian, we show that the magnetic interactions are both longer-ranged and more complex than was previously understood. The results provide the basis for accurate microscopic models of the finite temperature magnetic properties of Yttrium Iron Garnet, necessary for next-generation electronic devices.**

Yttrium Iron Garnet (YIG) is the 'miracle material' of microwave magnetics. Since its synthesis by Geller and Gilleo in 1957[4], it is widely acknowledged to have contributed more to the understanding of electronic spin-wave and magnon dynamics than any other substance[3]. YIG (chemical formula $Y_3Fe_5O_{12}$, crystal structure depicted in Fig. 1) is a ferrimagnetic insulating oxide with the lowest magnon damping of any known material. Its exceptionally narrow magnetic resonance linewidth — orders of magnitude lower than the best polycrystalline metals — allows magnon propagation to be observed over centimetre distances. This makes YIG both a superior model system for the experimental study of fundamental aspects of microwave magnetic dynamics[5] (and indeed, *general* wave and quasi-particle dynamics[6,7]), and an ideal platform for the development of microwave magnetic technologies, which have already resulted in the creation of the magnon transistor and the first magnon logic gate[5,8].

The unique properties of YIG have underpinned the recent emergence of new fields of research, including *magnonics*: the study of magnon dynamics in magnetic thin-films and nanostructures[5], and *magnon spintronics*, concerning structures and devices that involve the interconversion between electronic spin currents and magnon currents[1]. Such systems exploit the established toolbox of electron-based spintronics as well as the ability of magnons to be decoupled from their environment and efficiently manipulated both magnetically and electrically[5,9,10]. Significant interest has also developed in quantum aspects of magnon dynamics, using YIG as the basis for new solid-state quantum measurement and information processing technologies including cavity-based QED, optomagnonics, and optomechanics[11]. It has also recently been realised that one can stimulate strong coupling between the magnon modes of YIG and a superconducting qubit, potentially as a tool for quantum information technologies[12]. Spin

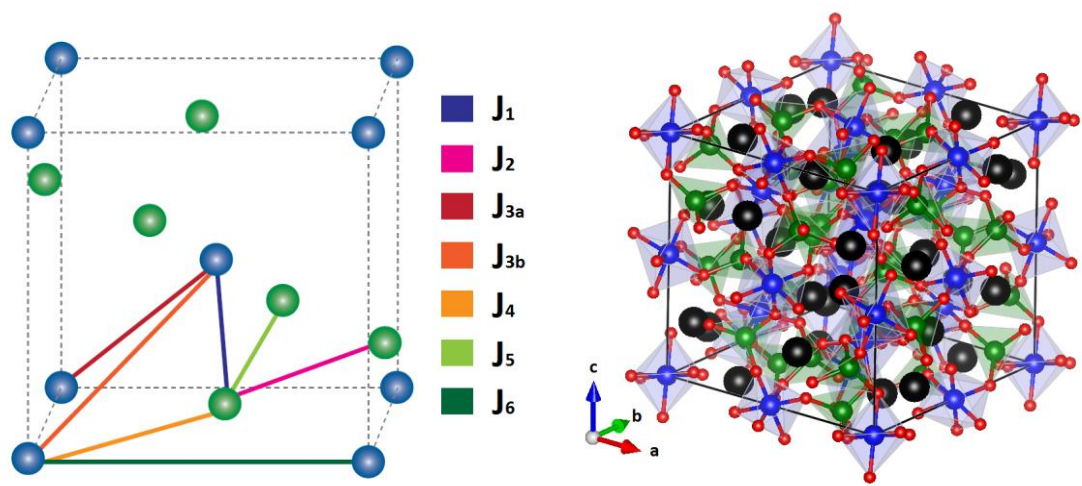

**Figure 1. Crystal structure and magnetic exchange paths in YIG.** *Left: First octant of the unit cell of YIG, indicating the two different $Fe^{3+}$ sites, with the tetrahedral sites in green and the octahedral sites in blue. Exchange pathways used in the Heisenberg effective Hamiltonian are labelled. Right: Unit cell of YIG, with the majority tetrahedral sites in green and the minority octahedral sites in blue. Black spheres are yttrium, red spheres are oxygen.*

1. Department of Physics, University of Oxford, Clarendon Laboratory, Oxford OX1 3PU, United Kingdom. 2. ISIS Facility, STFC Rutherford Appleton Laboratory, Harwell Campus, Didcot OX11 0QX, United Kingdom 3. Laboratory for Neutron Scattering and Imaging, Paul Scherrer Institut, CH-5232 Villigen, Switzerland. 4 INNOVENT e.V., Technologieentwicklung, Pruessingstrasse. 27B, D-07745 Jena, GERMANY caloritronics has also recently emerged as a potential application of YIG, utilising the spin Seebeck effect (SSE) and the spin Peltier effect (SPE) to interconvert between magnon and thermal currents, either for efficient large-scale energy harvesting, or the generation of spin currents using thermal gradients[13].

If the research into classical and quantum aspects of spin wave propagation in YIG is to achieve its potential, it is absolutely clear that the community requires the deep understanding of its mode structure, which only neutron scattering measurements can offer. In many theories and experiments, YIG is treated as a ferromagnet with a single, parabolic spin wave mode[14,15], simply because the influence of YIG's complex electronic and magnetic structure on spin transport is not known in sufficient detail. Such approaches must break down at high temperature when the optical modes are appreciably populated and a detailed knowledge of the structure of the optical modes is a necessary first step in any realistic model of the magnetic properties of YIG in this operational regime. Despite this, surprisingly little data exists relating to the detail of its magnon mode structure. The key previous work in this area is due to Plant et. al. [2], and dates back to the 70s. Using a triple-axis spectrometer, these early measurements were able to record 3 of the spin wave modes up to approximately 55 meV, but crucially there are 20 such modes and they are predicted to extend up to approximately 90meV (22 THz) [3].

Data were collected (see methods section) as a large, 4-dimensional hypervolume in frequency and momentum space, covering the complete magnon dispersion over a large number of Brillouin zones. Fig. 2 shows two-dimensional energy-momentum slices from this hypervolume with the wave-vector along three high-symmetry directions, normalised to a measurement on vanadium (see methods section). A large number of modes can be seen up to an energy of 80meV, whilst data in other slices show modes extending up to nearly 100 meV. The spectrum is dominated by a strongly dispersing and well-isolated acoustic mode at low energies (the so-called 'ferromagnetic' mode), and a strongly dispersing optical mode separated from this by a gap of approximately 30meV at the zone centre. Intersecting this upper mode is a large number of more weakly dispersing optical modes in the region of 30-50meV.

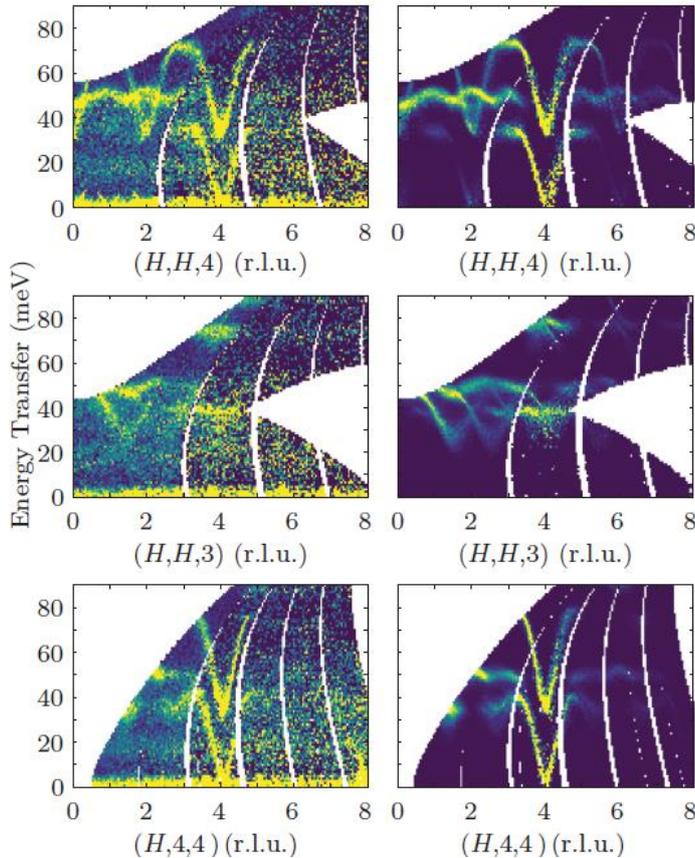

**Figure 2. Neutron scattering intensity maps of the magnetic excitation spectrum of YIG.** *a-c) Measured magnon spectrum along (H,H,4), (H,H,3), and (H,H,H) directions in reciprocal space, recorded in absolute units of mb sr$^{-1}$ meV$^{-1}$ f.u.$^{-1}$. d-f) resolution convoluted best fit to the model presented in the text, currently the basis for theoretical models of YIG. No scaling factors were used in the model.*

We model the data using a Heisenberg effective spin Hamiltonian, appropriate to YIG as it is both a good insulator and the $Fe^{3+}$ ions (S = $5/2$, g = 2) possess a negligible magnetic anisotropy due to the quenched orbital moment.

$$H = \sum_{i,j} \mathbf{S}_i^T J_{ij} \mathbf{S}_j + \sum_i \mathbf{S}_i^T A_i \mathbf{S}_i$$

We nevertheless include a magnetic anisotropy $A_i$ in our analysis to take into account crystal field effects, but find this term to be vanishingly small, consistent with previous results. The exchange matrix $J_{ij}$ is a general 3x3 matrix, whose elements are restricted by the symmetry of the bond connecting the spins $\mathbf{S}_i$ and $\mathbf{S}_j$. Following common practice, $J_{ij}$ is restricted to having only identical diagonal components (i.e. isotropic exchange) since anisotropic and off-diagonal contributions are likely to be small due to the lack of significant orbital angular momentum.

The spin Hamiltonian was diagonalized using the SpinW software package[16] and the calculated magnon dispersion was fitted by a constrained nonlinear least squares method to 1D cuts taken through the 2D intensity slices. We do not include any scaling factors for the magnon intensity, so the agreement between the model and the data in terms of absolute units is indicative of the quality of the model. Our final/best-fit model includes isotropic exchange interactions up to the 6$^{th}$ nearest neighbour, labelled $J_1$-$J_6$ in Fig 1. The exchanges in this work can be mapped to the exchanges commonly considered for YIG as follows: $J_{ad}$=$J_1$, $J_{dd}$=$J_2$, and $J_{aa}$ ={$J_{3a}$,$J_{3b}$}, where the subscript refers to the majority tetrahedral (d) and minority octahedral (a) sites. Due to the extremely large number of magnetic atoms (20) within the primitive cell, and the consideration of so many exchange pathways, this analysis would be impossible without the use of sophisticated software such as SpinW as the construction of an analytic model would be prohibitively time consuming. During the fitting process, features in the spectrum were weighted so that weak but meaningful features in the data were considered as significant as strong features. The SpinW model output is then convoluted with the calculated experimental resolution of the MAPS spectrometer, including all features of the neutron flight path and associated focussing/defocussing effects, as well as the detector coverage and effects from symmetrisation (see Supplementary Information for details). The final fitted values of the exchanges are listed in Table 1.

An important difference between our results and those of previous authors[2,3] is that there are two symmetry-distinct 3$^{rd}$-nearest-neighbour bonds (the so-called $J_{aa}$ in the literature, which we label $J_{3a}$ and $J_{3b}$) which have identical length but differ in symmetry. The $J_{3b}$ exchange lies precisely along the body-diagonal of the crystal, and thus has severely limited symmetry-allowed components owing to the high symmetry of the bond (point group $D_3$). The $J_{3a}$ exchange connects the same atoms with the same radial separation, but represents a different Fe-O-Fe exchange pathway as a result of the different point group symmetry (point group $C_2$), so it is distinguished from $J_{3b}$ by the environment around the Fe atoms. Models including anisotropic exchange or Dzyaloshinskii-Moriya interactions on the 1$^{st}$ -4$^{th}$ neighbour bonds were tested, but such interactions were found to destabilise the magnetic structure for arbitrarily small perturbations. We also find that $J_2$ ($J_{dd}$) is much smaller than previously supposed — the main effect of this exchange is to increase the bandwidth and split the optic modes clustered around 40 meV in a way contradicted by the data.

**Table 1**. *Fitted exchange parameters for YIG and their statistical uncertainties. The exchange constants are defined in Fig. 1*

| Exchange | This work (meV) | Ref. 3 (meV) |
|---|---|---|
| $J_1$ | 6.8(2) | 6.87 |
| $J_2$ | 0.52(4) | 2.3 |
| $J_{3a}$ | 0.0(1) | 0.65 |
| $J_{3b}$ | 1.1(3) | 0.65 |
| $J_4$ | -0.07(2) | - |
| $J_5$ | 0.47(8) | - |
| $J_6$ | -0.09(5) | - |

It has been pointed out[17] that the magnetic structure of YIG is incompatible with the cubic crystal symmetry, although to date no measurements have found any evidence for departures from the ideal cubic structure. Nevertheless, it is necessary to refine the magnetic structure in a trigonal space group (a symmetry that is experimentally observed in terbium rare earth garnets where the magnetoelastic coupling is much stronger[18]), in order to obtain a satisfactory goodness of fit, and a magnetic moment which agrees with bulk magnetometry[19]. Treating the unit cell of YIG in this fashion for the purposes of the SpinW simulation would be feasible, but would introduce a large number of free parameters which (given the very small size of the departure from cubic symmetry) would nevertheless be expected to change very little from a cubic model. This expectation is borne out by the excellent agreement between the data and the cubic spinwave model (see Fig. 2 and the supplementary materials for more details).

Features absent from the data which are relevant to technological applications (such as the conversion of microwave photons into magnons) include any strong indications of magnon-phonon or magnon-magnon coupling. The data are well described by a linear spinwave model, although the size of the 5$^{th}$-neighbour exchange is perhaps indicative of some small deviations not easily captured without such couplings. A strong magnon-phonon coupling would be expected to cause both broadening and anomalies in the dispersion of the magnon modes[20]. We do not observe any such effects, although our measurements would not be sensitive to any magnon-phonon coupling that shifts or broadens the spin wave signal by less than 3meV (the instrumental resolution).

Our results require a substantial revision of the impact of the optical modes on the room-temperature magnetic properties. The differences compared with the existing model are illustrated in Fig. 3, in which we plot the antisymmetric combination of transverse scattering functions $S_{xy}(Q,\omega) - S_{yx}(Q,\omega)$, which is proportional to the sign and magnitude of the measured spin-Seebeck effect arising from the associated magnon mode[21]. Most strikingly the absence of spectral weight in the flat mode at ~35meV, as well as a compression and shift of the 'positively' polarised (red) optical modes i.e. those modes which would precess counterclockwise with respect to an applied field. As has recently been shown, the thermal population, broadening, and softening of these modes at elevated temperatures substantially modifies the magnitude of the measured spin-Seebeck effect, which places limitations on device performance and determines the optimum operating temperature. Our results show that the distribution of optical modes is very different from what had previously been assumed, which has consequences for the temperature-dependent broadening[21]. Our new measurements can therefore be used as the basis for a precise microscopic model of the temperature dependent dynamical magnetic properties in YIG.

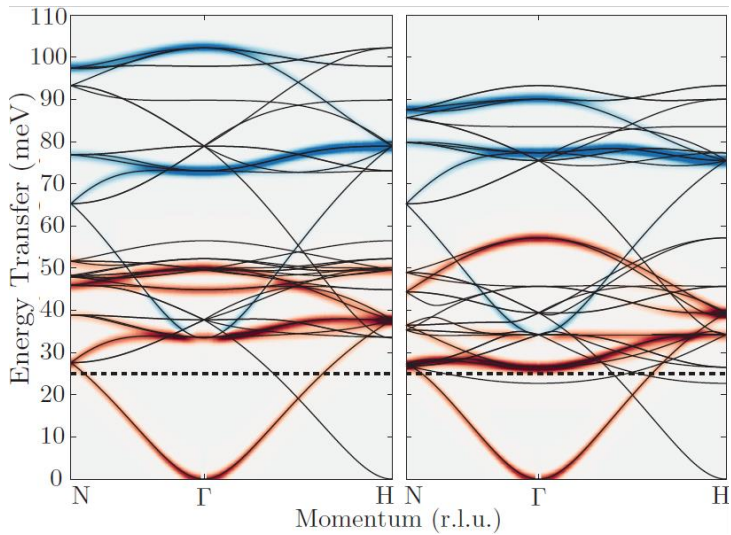

**Figure 3: Simulations of the magnon dispersion in YIG.** *(Right) the previous model[3,21] and (left) our new model of the magnon dispersion, where the colour and intensity correspond to the sign and magnitude of the correlator $S_{xy}(\mathbf{Q},\omega) - S_{yx}(\mathbf{Q},\omega)$, which is responsible for the spin Seebeck effect[21]. The horizontal line indicates $k_B T$ at room temperature.*

We have also estimated the parabolicity of the lowest lying 'ferromagnetic' magnon mode i.e. the point where the error of a quadratic fit becomes greater than 5%. We find this region to extend 14.8% of the way towards the Brillouin zone boundary along the H direction (the (0,0,1) direction in the centred unit cell), representing about 0.3% of the entire Brillouin zone. The departure from a purely parabolic acoustic magnon dispersion, as well as the population of optic magnon modes directly generates the temperature dependence of the spin Seebeck effect and our model can be used to fully understand such effects even at elevated temperatures through extension to a multi-magnon picture following the procedure in Refs. 21, and 22.

We have presented the most detailed and complete measurement of the magnon dispersion of YIG in a pristine, high quality crystal. Using linear spin-wave theory analysis we are able to reproduce the entire magnon spectrum across a large number of brillouin zones including a reproduction of the absolute intensities of the modes. We confirm the importance of the nearest-neighbour exchange, but are forced to radically reinterpret the nature and hierarchy of longer-ranged interactions. Our work has uncovered substantial discrepancies between previous models and the measured dispersion of the optical magnon modes in the 30-50 meV region, as well as the total magnon bandwidth, and the detailed nature of the magnetic exchanges. Through a detailed consideration of the symmetries of the exchange pathways and long-ranged interactions, we are able to fully reproduce the entire measured spectrum. Technological applications of YIG, particularly those utilising the spin Seebeck effect, are very sensitive to the optical magnons in the region of 30-40meV. This work overturns 40 years of established work on magnons in YIG and will be an essential tool for accurate modelling of the optical magnon modes in the room temperature regime.

**Acknowledgements**
This work was supported by the Engineering and Physical Sciences Research Council of the United Kingdom (grant nos. EP/J017124/1 and EP/M020517/1). We wish to acknowledge useful discussions with A. Karenowska, B. Hillebrandts, and T Hesjedal. We are grateful to S. Capelli (ISIS Facility) for use of the SXD instrument to characterise the crystals used in the experiments. Furthermore, C.D. would like to acknowledge the technical assistance of R. Meyer in the crystal growth of YIG. Experiments at the ISIS Pulsed Neutron and Muon Source were supported by a beamtime allocation from the United Kingdom Science and Technology Facilities Council.

**Author Contributions**
DP produced the preliminary optical floating zone crystal, and CD grew the flux crystal used in the final experiment. AJP and RAE performed the neutron experiments, with AJP responsible for the data reduction and RAE performed the instrumental resolution convolution. SW and AJP performed the data analysis and spinwave simulation, with advice and input from ST. AJP prepared the manuscript with input from all the co-authors. ATB supervised the project and assisted in the planning and editing of the manuscript.

## Methods

**Crystal Growth.** YIG crystal growth was carried out in high-temperature solutions applying the slow cooling method[23]. Starting compounds of yttrium oxide (99.999%) and iron oxide (99.8%) as solute and a boron oxide - lead oxide solvent were placed in a platinum crucible and melted in a tubular furnace to obtain a high-temperature solution[24]. Using an appropriate temperature gradient only a few single crystals nucleate spontaneously at the cooler crucible bottom and forced convection, obtained by accelerated crucible rotation technique (ACRT), allows a stable growth which results in nearly defect-free large YIG crystals[25]. The YIG crystal used in this study exhibits a size of 25 mm x 20 mm x 11 mm and a weight of 12 g. It was confirmed by neutron and X-ray diffraction that the YIG crystal was a single grain with a crystalline mosaic of approximately 0.07 degrees FWHM. Preliminary measurements were made on a crystal that was grown by the optical floating-zone method, starting from a pure powder of YIG.

**Neutron Scattering Data Collection and Reduction.**
Data were collected on the MAPS time-of-flight neutron spectrometer at the ISIS spallation neutron source at the STFC Rutherford Appleton Laboratory, UK. On direct geometry spectrometers such as MAPS, monochromatic pulses of neutrons are selected using a Fermi chopper with a suitably chosen phase. In our experiment neutrons with an incident energy ($E_i$) of 120 meV were used with the chopper spun at 350 Hz, giving energy resolution of 5.4 meV at the elastic line, 3.8 meV at an energy transfer of 50 meV, and 3.1 meV at an energy transfer of 90 meV. The spectra were normalized to the incoherent scattering from a standard vanadium sample measured with the same incident energy, enabling us to present the data in absolute units of mb sr$^{-1}$ meV$^{-1}$ f.u.$^{-1}$ (where f.u. refers to one formula unit of $Y_3Fe_5O_{12}$). Neutrons are scattered by the sample onto a large area detector on which their time of flight, and hence final energy, and position are recorded. The two spherical polar angles of each detector element, time of flight, and sample orientation allow the scattering function $S(\mathbf{Q}, \omega)$ to be mapped in a four dimensional space ($Q_x, Q_y, Q_z, E$). In our experiment the sample was oriented with the (HHL)-plane horizontal, while the angle of the (00L) direction with respect to the incident beam direction was varied over a 120 degree range in 0.25 degree steps. This resulted in coverage of a large number of Brillouin zones, which was essential in order to disentangle the 20 different magnon modes. Due to the complex structure factor resulting from the number of Fe atoms in the unit cell, the mode intensity varies considerably throughout reciprocal space. The large datasets recording the 4D space of $S(\mathbf{Q},\omega)$, ~100 GB in this case, were reduced using the Mantid framework[26], and both visualized and analyzed using the Horace software package[27]. Taking advantage of the cubic symmetry of YIG, the data were folded into a single octant of reciprocal space (H>0, K>0, L>0), adding together data points that are equivalent in order to produce a better signal-to-noise. 2D slices were taken from the 6 reciprocal space directions depicted in Figure 2 and in Supplementary Information.

# Supplementary Information For: The Final Chapter In The Saga Of YIG

A. J. Princep[1]*, R. A. Ewings[2], S. Ward[3], S. Tóth[3], C. Dubs[4], D. Prabhakaran[1], A. T. Boothroyd[1]

1. Department of Physics, University of Oxford, Clarendon Laboratory, Oxford OX1 3PU, United Kingdom.
2. ISIS Facility, STFC Rutherford Appleton Laboratory, Harwell Campus, Didcot OX11 0QX, United Kingdom
3. Laboratory for Neutron Scattering and Imaging, Paul Scherrer Institut, CH-5232 Villigen, Switzerland.
4 INNOVENT e.V., Technologieentwicklung, Pruessingstrasse. 27B, D-07745 Jena, GERMANY

**Contents:**

1) Additional information on the fitting procedure

2) Symmetry of the 3rd neighbour Exchange

3) Additional data slices simulated, and detailed comparison with previous model.

4) Instrumental broadening and simulation

5) Supplementary References

1. **Additional information on the fitting procedure**

To extract the exchange parameters fitting was performed by a bounded non-linear least squares fit to the following cuts of the experimental data set:

| Q-BASIS | Q RANGE (R.L.U) | ENERGY RANGE (MEV) |
|---|---|---|
| H H 4 | 0.0 0.05 3.5 | 40 1.5 80 |
|  | 2.5 0.05 5.0 | 08 1.5 90 |
| H H H | 0.0 0.05 3.5 | 20 1.5 60 |
|  | 3.0 0.05 5.0 | 08 1.5 90 |
| H H 3 | 0.0 0.05 5.0 | 09 1.5 90 |
| 2 2 L | 0.0 0.05 1.0 | 40 1.5 50 |
|  | 3.0 0.05 5.0 | 20 1.5 90 |
| 3 3 L | 0.0 0.05 5.0 | 09 1.5 90 |

Initial fitting was performed on exchanges $J_2$, $J_{3a}$, $J_{3b}$, $J_4$, $J_5$ and $J_6$ where the results were subsequently used as starting points for when the exchanges $J_1$, and the anisotropy parameter D were allowed to vary. Due to the exceptional quality of the data, the background signal was approximated as a constant value, which was unique to each cut. As well as this, a common intensity factor and convolution width was used for all cuts. The convolution width was not fitted in the initial procedure, rather it was fitted separately to a 1D cut from (H, H, 4) with integration 1.9-2.1 r.l.u and energy binned between 9.0 and 90 meV in steps of 1.5 meV. The ferromagnetic features around (4, 4, 4) were found to be dominant over the intricate higher energy features in the basic fitting approach. Masking this feature led to unsatisfactory parameter convergence, so to overcome this a weighting factor was introduced, which allowed a parameter convergence describing both low and high energy features.

During the fitting procedure parameters were allowed to vary between their boundary conditions:

| Exchange | Value | Boundary condition |
|---|---|---|
| $J_1$ | 6.7600 (0.2025) | [3 10] |
| $J_2$ | 0.5207 (0.0370) | [0 1] |
| $J_{3a}$ | 0.0001 (0.1362) | [-2 2] |
| $J_{3b}$ | 1.0539 (0.3216) | [-2 2] |
| $J_4$ | -0.0686 (0.0152 ) | [-1 1] |
| $J_5$ | 0.4736 (0.0840) | [0 1] |
| $J_6$ | -0.0930  (0.0499) | [-1 1] |
| D | 0.01000  (0) | [-1 1] |

Starting parameters were randomly selected within the limits and were free to vary within the boundary conditions. It was found that only those around the presumed starting parameters gave a meaningful convergence.

## 2. Symmetry of the 3rd neighbour Exchange

The difference between $J_{3a}$ and $J_{3b}$ can be understood easily from a projection of the crystal structure along the axis of the bond, depicted in figure S1.

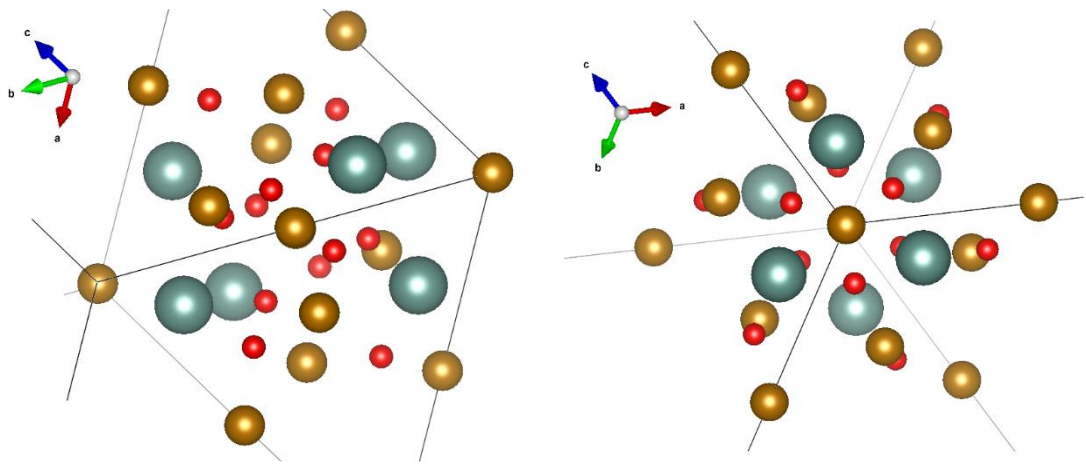

**Figure S1.** The left image depicts the projection of the crystal structure along the J3a bond, from which the 2-fold symmetry can be seen. The right image depicts the projection of the crystal structure along the J3b bond, from which the symmetry can be seen to obey the higher symmetry D3 point group. These images were generated using VESTA[S1].

### 3. Additional data slices simulated, and detailed comparison with previous model.

Further 2D slices to the data that were used for fitting are found below in figure S2. Figure S3 depicts a detailed comparison of the optical magnon modes in the region 30-50meV, contrasting the model presented in this work with that of previous authors[3].

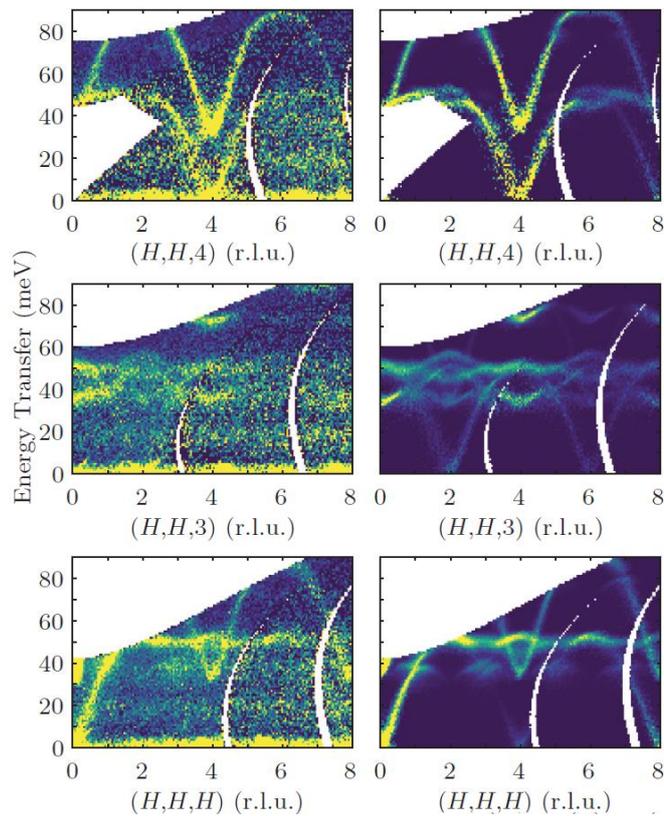

**Figure S2**. Supplementary neutron scattering intensity maps along (H,H,4), (H,H,3), and (H,H,H) directions in reciprocal space. Left: Experimental data. Right: Model.

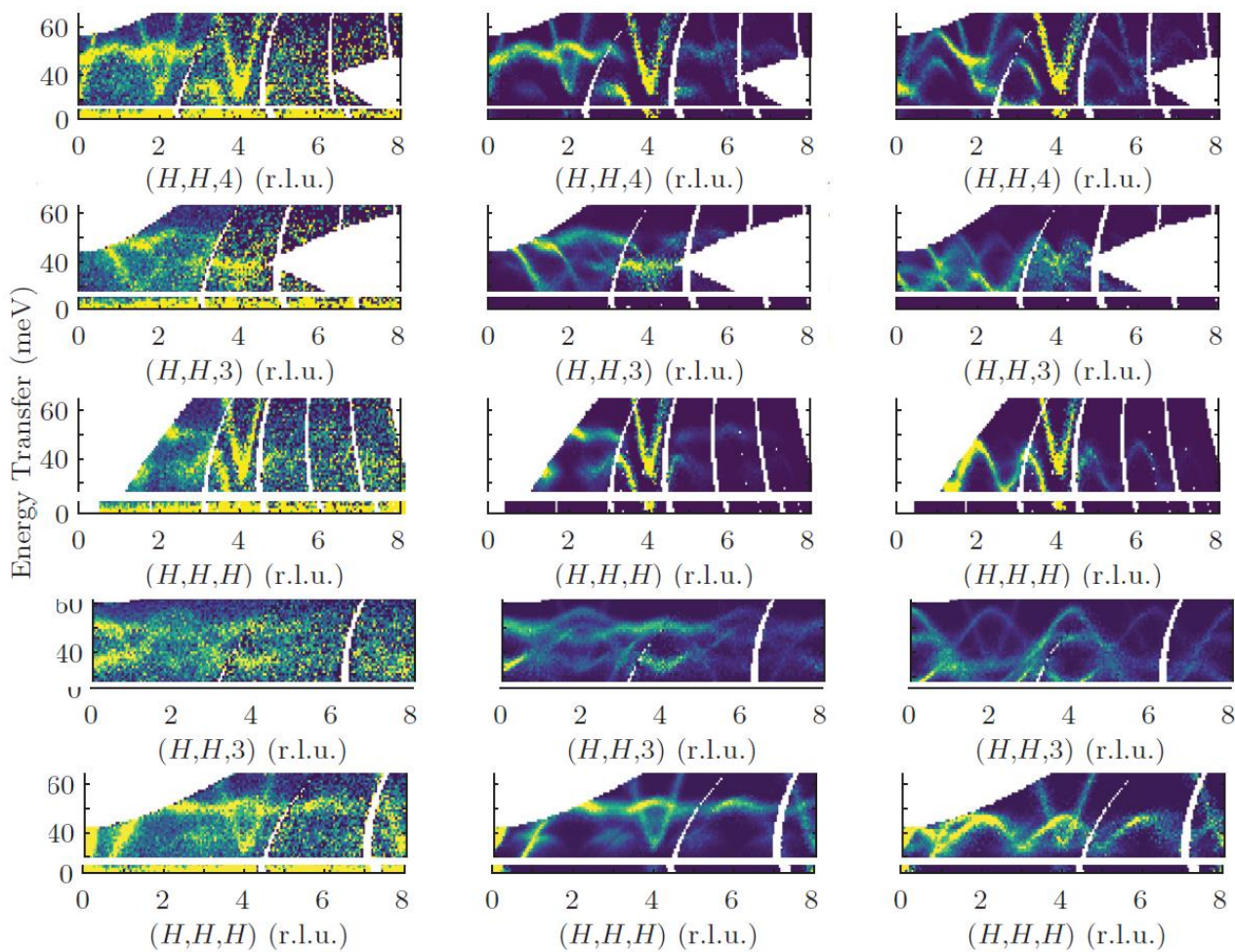

**Figure S3.** Zoomed figures emphasising comparison between the model in this paper (middle column) and the model owing to previous work[3] (right column). The left column shows the corresponding data in the same region of reciprocal space.

## 4. Instrumental broadening and simulation

The broadening of the signal due to instrumental resolution arises due to several effects. The main contributions are:
- Deviations in neutron arrival time (at both the chopper and detectors) due to the finite depth of the source moderator and the angle of the moderator face with respect to the beamline collimation, and hence deviations in the departure time of the neutron burst compared to the notional zero time.
- Deviations in neutron arrival time due to the finite sweep time, length and slit width of the monochromating Fermi chopper.
- The divergence of the scattered beam due to the finite size of the sample subtended at each detector element.
- The divergence of the incident neutron beam due to the finite size of the beam and the geometry of the beamline collimation

These effects are computed in the reference frame of the spectrometer and then converted into the reference frame of the sample and convoluted. The convolution is performed by Monte-Carlo sampling over the resolution width of each of the terms described above. In general each **Q**-energy bin in the images shown in Fig. 2 of the main article (as well as supplementary figures S2 and S3) contains data from many detector elements taken with many different sample orientations. The resolution for each element was accounted for in the corresponding simulations shown in the same figures.

## 4. Supplementary references

[S1] K. Momma & F. Izumi. VESTA: a three-dimensional visualization system for electronic and structural analysis. *J. Appl. Crystallogr.*, 41, 653-658 (2008)